# The transport, ultrasound, and structural properties for the charge ordered $Pr_{1-x}Ca_xMnO_3$ ($0.5 \le x \le 0.875$) manganites

R. K. Zheng, G. Li, Y. Yang, A. N. Tang, W. Wang, T. Qian, X. G. Li [a)]

Structure Research Laboratory, Department of Materials Science and Engineering,

University of Science and Technology of China, Anhui, Hefei 230026, P. R. China

The effects of the cooperative Jahn-Teller effect on the crystal structure and the stability of the charge ordered (CO) state were studied by measurements of powder X-ray diffraction, resistivity, and ultrasound for $Pr_{1-x}Ca_xMnO_3$ ($0.5 \le x \le 0.875$). Powder X-ray diffraction revealed a change of the crystal structure from tetragonally compressed to tetragonally elongated orthorhombic between $x$=0.75 and $x$=0.8 in the CO state, resulting from the crossover of the cooperative Jahn-Teller vibration mode from $Q_2$ to $Q_3$. The relative stiffening of the ultrasound ($\Delta V/V$) reflecting the magnitude of the cooperative Jahn-Teller lattice distortion in the CO state increases with increasing $x$ from 0.5 to 0.625, reaching the largest and being almost $x$-independence for $0.625 \le x \le 0.8$, and drops steeply with further increase of $x$. Coincident with the variation of the $\Delta V/V$ with $x$, the stability of the CO state reflected by the magnetoresistance effect increases with increasing $x$ from 0.5 to 0.625, reaching the most stable for $0.625 < x < 0.825$, and becomes unstable with further increase of $x$. These features demonstrate that the cooperative Jahn-Teller lattice distortion is one of the key ingredients in understanding the essential physics of the CO state in manganites.

[a)] Corresponding author, Electronic mail: lixg@ustc.edu.cn

# I. Introduction

Hole-doped perovskite type manganites $R_{1-x}A_xMnO_3$ (where R and A are trivalent rare-earth and divalent alkaline-earth ions, respectively) are highly correlated electron systems in which the strong interplay among the spin, charge, lattice, and orbital degrees of freedom leads to complex electronic, magnetic, and structural phase diagrams and many dramatic phenomena such as the colossal magnetoresistance (CMR) effect, the charge ordering, the phase separation [1-3]. The $Pr_{1-x}Ca_xMnO_3$ series are typical examples of charge ordering cases appeared in a wide range of doping levels $0.3 \le x < 0.875$ [4]. At the low doping levels of $0.3 \le x \le 0.5$, the $Pr_{1-x}Ca_xMnO_3$ series shows notorious concomitant charge and orbital ordering (OO) with distinct changes of the lattice parameters around the charge ordering transition temperature $T_{CO}$. The long-range antiferromagnetic ordering, however, establishes at temperatures much lower than $T_{CO}$ [4]. Numerous experimental results have shown that the charge ordered (CO) state in low-doped ($0.3 \le x \le 0.5$) $Pr_{1-x}Ca_xMnO_3$ series can be easily destroyed by application of external magnetic field [5], hydrostatic pressure [6], electric field [7,8], X-ray illumination [9], oxygen isotope exchange [10], and A- or B-site ion substitution [11,12]. The insulator-to-metal transition induced by illumination with X-ray, application of hydrostatic pressure, and oxygen isotope exchange for charge ordered $Pr_{0.7}Ca_{0.3}MnO_3$ [9], $(La_{0.25}Nd_{0.75})_{0.7}Ca_{0.3}MnO_3$ [6], $(La_{0.175}Pr_{0.525})_{0.7}Ca_{0.3}MnO_3$ [10], respectively, undoubtedly manifest the strong interactions between the conduction electrons and the cooperative Jahn-Teller (JT) lattice distortions associated with the Mn site. Our recent systematical studies on the role of the cooperative JT effect for analogous charge ordered $La_{1-x}Ca_xMnO_3$ ($0.5 \le x \le 0.9$) series showed that the crystal and magnetic structures, the orbital ordering, the phase separation tendency, the strength of the ferromagnetism, the magnitude of the lattice distortion and the stability of the CO state are closely dependent on the cooperative JT effect with different vibration modes [13,14], which evidence that the cooperative JT effect plays a key role in understanding the rich physics of the CO state. Although intense theoretical and experimental studies have been devoted to the CO state for $Pr_{1-x}Ca_xMnO_3$ series, most of the studies, however, focus their attention on the low-doped ($x$≤0.5) series of this system [5,7,8,15-17]. A deep understanding on the underlying physics of the CO state, especially the potential role played by the cooperative JT effect, for Ca-high-doped (0.5≤$x$≤0.875) $Pr_{1-x}Ca_xMnO_3$ series is yet to be explored.

The ultrasound measurement is a sensitive tool in studying the cooperative JT phase transition and has been successfully employed to study the static/dynamic JT effect for perovskite type manganites such as $Pr_{1-x}Ca_xMnO_3$ ($x$=0.35, 0.4, 0.5) [17,18] and $La_{1-x}Ca_xMnO_3$ (0.5≤$x$≤0.9) [13,19,20]. In this paper, we studied the role of the cooperative JT effect in the CO state using ultrasound. The influences of the cooperative JT effect on the stability of the CO state and the crystal structure were the particular interesting. We found that the magnitude of the magnetoresistance (*MR*) effect representing the stability of the CO state strongly depends on the magnitude of the cooperative JT lattice distortion. The crystal and magnetic structures in the CO state were also found to be related to different types of JT vibration mode. All these effects suggest that the cooperative JT effect is the dominant physics in the CO state for the charge ordered $Pr_{1-x}Ca_xMnO_3$ (0.5≤$x$≤0.875).

## II. Experimental details

The polycrystalline $Pr_{1-x}Ca_xMnO_3$ ($x$=0.5, 0.55, 0.6, 0.625, 0.67, 0.7, 0.75, 0.8, 0.825, 0.85, 0.875) series were synthesized by a standard solid-state ceramic technique. Stoichiometric amounts of high-purity $Pr_6O_{11}$, $CaCO_3$, and $MnCO_3$ were weighed out and intimately mixed in an agate mortar and pestle. The reactants were fired in air at 1050ºC, 1150ºC, and 1350ºC, respectively, for 20 hours with regrinding between firings. Finally, the fired powder were pelleted, and fired at temperatures varying from 1360ºC to 1420ºC for 20 hours according to the doping level $x$.

Resistivity $\rho$(T) was measured using a standard four-probe technique at temperatures from 4.2 K to 300 K in magnetic fields up to 14T. Powder X-ray diffraction measurements were made on the MXP18AHF powder X-ray diffractometer (MAC Science Co. Ltd., Japan) using Cu $K\alpha$ radiation ($\lambda$=1.54056 Å) at temperatures from 40 K to 300 K. The longitudinal ultrasound was measured using the digital AUW-100 Advanced Ultrasonic Workstation (Matec Instrument Companies, USA) at temperatures from 20 K to 320 K. A 15 MHz $LiNbO_3$ transducer was used for the generation of the longitudinal ultrasound. Sound velocity ($V$) was calculated through the following relationship :

$$V = 2 \times L / t$$

where $L$ is the thickness of the sample, $t$ is the time difference between two successive pulse echoes. The relative change of the sound velocity ($\Delta V/V$) was calculated according to the following relationship :

$$\Delta V / V = \frac{V_{\max} - V_{\min}}{V_{\min}} = \frac{V_{\max} - V_{T_{CO}}}{V_{T_{CO}}}$$

where the $V_{max}$ is the maximal sound velocity below $T_{CO}$, the $V_{min}$ (=$V_{T_{CO}}$) is the minimum sound velocity in the temperature range from 20 K to 320 K. All data were collected upon warming.

## III. Results and discussion

Fig. 1 shows the temperature dependence of the resistivity $\rho$(T) under zero and external fields ($H$) up to 14T for the $Pr_{1-x}Ca_xMnO_3$ series. For the $x$=0.5 compound, upon cooling from room temperature the zero-field resistivity curve shows semiconductive behavior with a discernible change in the slope around $T_{CO}$, implying the onset of the charge ordering phase transition. With increasing fields (especially for $H$>9T) the resistivity below $T_{CO}$ was drastically reduced, exhibiting a large CMR effect in the CO state. Particularly, under a field of 14T, the semiconductive behavior of the resistivity below $T_{CO}$ for $H$=0T was suppressed, and instead, the resistivity shows an insulator-metal transition around $T_P$=175 K, as indicated by the arrow shown in Fig. 1. The large decrease of the resistivity and the downward shift of $T_{CO}$ by about 110 K upon the application of a field of 14T manifest the unstable nature of the CO state for the $x$=0.5 compound. This field induced "melting" of the CO state and the unusual resistivity behavior for $H$=14T can be understood within the framework of the phase separation model, which will be discussed later. Similar to that observed in the $x$=0.5 compound, the resistivity for $x$=0.55 was also drastically suppressed when a field of 14T is applied, as seen in Fig. 1. Nevertheless, a field of 14T is not large enough to induce an insulator-metal transition as observed in the $x$=0.5 compound. The resistivity still remains to be semiconductive persisting down to 90 K, and no sign of insulator-metal transition is found, which indicates that the stability of the CO state becomes stronger as the Ca concentration $x$ increases from 0.5 to 0.55. We note that with further increase of the Ca concentration $x$ from 0.55 a field of 14T becomes much less effective on the resistivity, suggesting the stability of the CO state continues to increase. From $x$=0.625 to 0.825, the responses of the resistivity at low temperature ($T \ll T_{CO}$) to the application of a field of 14T are almost neglectable. When the Ca concentration $x$ is larger than $x$=0.825, the resistivity becomes increasingly field dependence. These magnetic field effects on the resistivity (or the stability of the CO state) are, in generally, similar to those observed in the Ca-high-doped charge ordered $La_{1-x}Ca_xMnO_3$ analogues [13].

Based on above magnetic field effects on the behavior of the resistivity, we here focus on our attention on the electronic conduction mechanism in the CO state and the variation of the stability of the CO state with the doping level $x$. First, we make an analysis of the electronic conduction mechanism in the CO state. Previous studies indicated that the electronic conducting mechanism in the CO state for perovskite-type manganites can be best described by variable-range hopping (VRH) model [21]. The $\rho$(T) behavior due to variable-range hopping in three-dimensional (3D) system can be best fitted by the law [22] : $\rho = \rho_0 \exp(T_0 / T)^{1/4}$. The parameter $T_0$ is a characteristic temperature, and is related to the density of states $N(E_F)$ at the Fermi level and the charge carrier localization length. We fit the $\rho$(T) data using the VRH model. Fig. 2(a) shows the $\ln \rho$ versus $T^{-1/4}$ curves for $Pr_{0.5}Ca_{0.5}MnO_3$ at various magnetic fields. It is seen that the zero-field resistivity for T≤160 K can be well fitted by $\rho = \rho_0 \exp(T_0 / T)^{1/4}$, which indicate that VRH is the dominant electronic conduction mechanism in the charge ordered phase. It is interesting to find that with increasing field from 0T to 9T the highest temperature, here defined as $T_{VRH}$, where linear $\ln \rho$ -$T^{-1/4}$ relationship was found, decreases from $T_{VRH}$=160 K at 0T to $T_{VRH}$=106 K at 9T. The field dependence of $T_{VRH}$ is shown in the inset (b) of Fig. 2(a). The decrease of the $T_{VRH}$ with increasing field implies that the electronic conduction mechanism in the CO state gradually deviates from the VRH model, and other conduction mechanism plays more and more important role. Using a dark-field image technique, Mori *et al.* [23] directly observed the nano-size ferromagnetic (FM) metallic clusters embedded in the charge ordered antiferromagnetic (AFM) insulator matrix (i.e. phase separation) at temperatures much lower than $T_{CO}$ in analogous $La_{0.5}Ca_{0.5}MnO_3$ compound. For our $Pr_{0.5}Ca_{0.5}MnO_3$ case, the $\rho - T$ behaviors and the deviation of the electronic conduction mechanism from VRH with increasing field can be explained by the phase separation model. For *H*=0T, the CO/AFM phase dominates the physics. The electronic hopping is, thus, governed by the VRH. The application of an external field induces the FM clusters growing in size and fraction, leading to distinct coexistence of FM metallic and CO/AFM phases, which is particularly prominent under a high field. Part of the electrons, therefore, conduct through the FM metallic phases. Two electronic conduction channels (i.e. FM/metallic and CO/AFM) both contribute to the conduction of the electrons. A deviation of the electronic conduction from VRH is, thus, naturally expected with increasing field.

We found that, under a high field of 14T, no linear $\ln \rho - T^{-1/4}$ relationship can be found at any temperature region for $T<T_{CO}$. The application of a field of 14T induces an insulator-metal transition around $T_P$=175K, as seen in the inset (a) of Fig. 2(a). However, the insulator-metal transition is incomplete. When the temperature is lowered to $T_{CO}$=130 K the resistivity increases sharply. This $\rho - T$ behavior, in fact, can also be understood within the framework of the phase separation model. Under a high field the FM metallic and CO/AFM clusters coexist and constitute a parallel conduction [24,25], the resistivity, therefore, can be described by the following equation [25] :

$$\frac{1}{\rho}=\frac{1}{\rho_{FM}}+\frac{1}{\rho_{CO}}=\frac{1}{A+BT^{2.5}}+\frac{1}{\rho_0 \exp(T_0/T)^{1/4}} \quad (1)$$

The term $\rho_{FM}=A+BT^{2.5}$ is an empirical equation used to describe the electronic conduction in the FM metallic clusters [25]. The A and B correspond respectively to the scattering by defects, and by a combination of phonons, electrons, and spin fluctuations [25]. The term $\rho_{CO}=\rho_0 \exp(T_0/T)^{1/4}$ is used to describe the electronic conduction in the CO/AFM clusters. Using Eq.(1), we fitted the $\rho(T)$ data at $H$=14T for T≥$T_{CO}$. The fit is satisfactory. Here, it's interesting to compare the resistivity behaviors under high fields for $Pr_{0.5}Ca_{0.5}MnO_3$ with that for $La_{0.5}Ca_{0.5}MnO_3$ analogue. In Fig. 2(b), we show the temperature dependence of the resistivity under various fields for $La_{0.5}Ca_{0.5}MnO_3$. Similar to that observed in $Pr_{0.5}Ca_{0.5}MnO_3$, the resistivity data in the CO state (T≤$T_{VRH}$) for $H$≤10T can also be well fitted using equation $\rho_{CO}=\rho_0 \exp(T_0/T)^{1/4}$, hinting the VRH dominates the electronic conduction mechanism in the CO state. Like that for $Pr_{0.5}Ca_{0.5}MnO_3$, the $T_{VRH}$ for $La_{0.5}Ca_{0.5}MnO_3$ also decreases with increasing field, which can also be explained based on the phase separation model mentioned above. We note that the application of a field of 12T induces an incomplete insulator-metal transition with the peak position around 135K (inset (a) in Fig. 2(b)), which is, in fact, similar to the resistivity behavior under 14T for $Pr_{0.5}Ca_{0.5}MnO_3$ (inset (a) in Fig. 2(a)). The similarity in the resistivity between them implies the same underlying mechanism of physics. What is different from $Pr_{0.5}Ca_{0.5}MnO_3$ is that a field of 14T is sufficient large to induce a complete insulator-metal transition around 150K for $La_{0.5}Ca_{0.5}MnO_3$, which suggest the CO state for $Pr_{0.5}Ca_{0.5}MnO_3$ is more stable than that for $La_{0.5}Ca_{0.5}MnO_3$. Note that both the $Pr_{0.5}Ca_{0.5}MnO_3$ and $La_{0.5}Ca_{0.5}MnO_3$ have the CE-type magnetic structure, $3d_{x^2-r^2}/3d_{y^2-r^2}$ orbital ordering, *b*-axis tetragonally compressed orthorhombic crystal structure [26-28], and the same $Mn^{3+}$ : $Mn^{4+}$ ratio. The only difference between them is the average A-site ionic radius. Since the ionic radius for $Pr^{3+}$ ion ($r_{Pr^{3+}} = 1.179$ Å) is smaller than that for $La^{3+}$ ion ($r_{La^{3+}} = 1.216$ Å) [29], the buckling of the Mn-O-Mn bond for $Pr_{0.5}Ca_{0.5}MnO_3$ is, therefore, more prominent than that for $La_{0.5}Ca_{0.5}MnO_3$, which, as a result, enhance the localization character of the itinerant $e_g$ electrons, and hence a more stable CO state for $Pr_{0.5}Ca_{0.5}MnO_3$.

To find the electronic conduction mechanism under high fields for $La_{0.5}Ca_{0.5}MnO_3$ and compare with that for $Pr_{0.5}Ca_{0.5}MnO_3$, we fitted the $\rho(T)$ data at *H*=12, 13, and 14T for $La_{0.5}Ca_{0.5}MnO_3$ using Eq.(1). The results are shown in the inset (a) of Fig. 2(b). It is seen that the fit is satisfactory, especially for *H*=13 and 14T. This feature, together with the phase separation experimentally observed in $La_{0.5}Ca_{0.5}MnO_3$ by Mori *et al.* [23], strongly demonstrate that there exists two parallel electronic conduction channels under a high field for $La_{0.5}Ca_{0.5}MnO_3$, which, in turn, also confirm that the phase separation and parallel electronic conduction channels exist under a high field in $Pr_{0.5}Ca_{0.5}MnO_3$.

To complete the characterization of the electronic conduction in the CO state, we show the $\ln\rho - T^{-1/4}$ curves at *H*=0T for $Pr_{1-x}Ca_xMnO_3$ (*x*=0.55, 0.6, 0.625, 0.67, 0.7, 0.75, 0.8, 0.825) in Fig. 3. It is straightforward to find that the electronic conduction in the CO state for the serial compounds can be best described by the VRH mechanism. However, the conduction mechanism starts to deviates from VRH for *x*>0.75, evidenced by the poorer linearity of the $\ln\rho - T^{-1/4}$ curves at low temperatures. This may be due to the fact that the formation of the charge ordering tendency and hence the fraction of the CO/AFM phase decreases because of the decrease of the concentration of the $Mn^{3+}$ ions. Although we do not have the direct evidence for the decrease of the fraction of the CO/AFM phase for *x*>0.75, it was found that the charge ordering in analogous $La_{1-x}Ca_xMnO_3$ compounds become increasingly weaker with increasing $x$ from 0.75, which is evidenced by the weaker and broader superlattice peaks observed in the electron diffraction patterns [2].

Now, Let's turn the attention to the variation of the stability of the CO state with the doping level $x$ and its correlation with the JT effect. Although a qualitative analysis of the stability of the CO state is made above, to quantitatively characterize the stability of the CO state, we calculated the $MR(MR = (\rho_0 - \rho_H)/\rho_H)$ effect at the field of 14T and the temperature of 90K for $Pr_{1-x}Ca_xMnO_3$ series. For discussion convenience, we present the $x$-dependence of the $MR$ at 14T and 90K in Fig. 5. It is seen that the $MR$ effect is the largest for $x$=0.5 and is the order of $10^4$, and drops rapidly with the increase of $x$ from 0.5 to 0.625. We note that the $MR$ effect reaches its minimum and almost $x$-independence for $0.625<x\leq0.825$. For $x>0.825$, the $MR$ effect increases with increasing $x$. The variation of $MR$ effect with $x$ strongly reflects that the CO state is unstable for $x$=0.5, nevertheless, becomes increasingly stable with increasing $x$ from 0.5 to 0.625, and then is the most stable for $0.625<x\leq0.825$. For $x>0.825$, the CO state becomes sensitive to external field again. As is well known, the strong electron-lattice coupling via the Jahn-Teller distortion favors the localization of charge carriers, while the spin-charge coupling via the DE induced ferromagnetism demands the active hopping of charge carriers. It is the competition between these two couplings (*i.e.* the electron-lattice and spin-charge via DE) that is mainly responsible for the charge transport properties and the stability of CO state. An external magnetic field tends to align electron spins along its direction and thus results in the increase of the effective strength of DE interaction, which in turn enhances the FM coupling between neighboring Mn spins as well as the mobility of charge carriers, increasing the total size of the FM domains and reducing that of the insulating CO domains. As mentioned above, there exists obvious phase separation in the $Pr_{0.5}Ca_{0.5}MnO_3$ compounds. This phase separation results in the partial "melting" of the CO state (see Fig. 1) via enhanced DE interaction under external field. However, with increasing $x$ from 0.5 the phase separation and ferromagnetism of the system is suppressed due to the increase of the Jahn-Teller distortion which will enhance the stability of the CO state.

Our recent systematic studies on the charge ordered $La_{1-x}Ca_xMnO_3$ ($0.5 \leq x \leq 0.9$) series shown that the magnitude of the *MR* effect representing the stability of the CO state is closely dependent on the magnitude of the cooperative JT lattice distortion in the CO state [13,30]. However, the relationship between the stability of the CO state and the cooperative JT lattice distortion in the present $Pr_{1-x}Ca_xMnO_3$ ($0.5 \leq x \leq 0.875$) series is still unknown. To probe into this issue, we studied the ultrasonic responses to the cooperative JT phase transition. In Fig. 4, we show the temperature dependence of the relative changes of the ultrasound ($\Delta V/V$) for the $Pr_{1-x}Ca_xMnO_3$ series. Similar to that observed in the charge ordered $La_{1-x}Ca_xMnO_3$ ($0.5 \leq x \leq 0.87$) series, the ultrasound softens considerably as the temperature is lowered from 300 K, reaches its minimum at $T_{CO}$, and then stiffens drastically with further cooling of the temperature. These behaviors of the ultrasound, in fact, are fundamental features of the cooperative JT phase transition [31]. The stiffening of the ultrasound below $T_{CO}$, therefore, is an indication of the development of the cooperative JT effect. The relative stiffening of the ultrasound ($\Delta V/V$) can be viewed as a scale of the magnitude of the cooperative JT distortion in the CO state. We present the $\Delta V/V$ as a function of the Ca concentration $x$ in Fig. 5. It is seen that, with increasing $x$ from 0.5 to 0.625, the $\Delta V/V$ increases rapidly, and becomes almost saturated and $x$-independence for $0.625 \leq x \leq 0.8$. For $x > 0.8$, the $\Delta V/V$ drops drastically with increasing $x$. The $\Delta V/V \sim x$ relationship reflects a variation of the magnitude of the cooperative JT lattice distortion with $x$ in the CO state. From Fig. 5, one can intuitively find that the *MR* decreases rapidly with increasing $\Delta V/V$, and reaches the smallest and being almost $x$-independence for $0.625 \leq x \leq 0.8$ where the $\Delta V/V$ is the largest. As for $x > 0.825$, the *MR* starts to increase with the increase of $x$, consistent with the decrease of $\Delta V/V$. Based on the strong correlation between the $\Delta V/V$ and the *MR* for the present $Pr_{1-x}Ca_xMnO_3$ series and the previous studied $La_{1-x}Ca_xMnO_3$ series [13,14], it seemed that the cooperative JT lattice distortion and resultant electron-lattice interaction is the key factor that determines the stability of the CO sate for $R_{1-x}Ca_xMnO_3$ (R=La, Pr) ($0.5 \leq x \leq 0.875$).

To further look into the cooperative JT effect and its influence on the crystal structure in the CO state, we carried out a series of low temperature powder X-ray diffraction measurements. We analyzed the obtained powder X-ray diffraction patterns using the Rietveld refinement method [32] based on the space group *Pnma*. Typical powder X-ray diffraction patterns and the refinement results at 40 K for the $x$=0.625 and 0.8 compounds are shown in Fig. 6. The reliable parameter $R_P$ at temperatures from 40 K to 300 K lies between 10-15% for the compounds, which implies that the refinement results are acceptable. Fig. 7 shows the variation of the orthorhombic lattice parameters $a$, $b/\sqrt{2}$, and $c$ with temperature for selected $x$=0.625, 0.67, 0.75, and 0.825 compounds. It is seen that with decreasing temperature from 300 K the lattice parameters changes drastically near $T_{CO}$. The large changes of the lattice parameters near $T_{CO}$ are obviously mainly due to the development of the cooperative JT lattice distortion (i.e. orbital ordering) [3,33]. In fact, at high temperatures ($T>T_{CO}$), the local JT distortions are randomly oriented along the crystallographic axes equivalently and the macroscopic symmetry of the lattice is an average over the local distortion. When the temperature is cooled below $T_{CO}$, the configurational interactions between local distortions cause them to align themselves cooperatively along a single axis, and thus leading to a macroscopic distortion of the entire crystal [31]. We note that Hazama *et al*'s [17] recent observation of the softening of the elastic constant of $(C_{11}-C_{12})/2$ above $T_{CO}$ in $Pr_{1-x}Ca_xMnO_3$ ($x$=0.35, 0.4, 0.5) single crystals strongly indicates the coupling of the quadrupole moment of the $e_g$ orbital to the elastic strains, leading to distinct structural changes associated with quadrupole (orbital) ordering. On the other hand, it was also observed that the coupling of the charge-fluctuation mode of the $Mn^{3+}$ and $Mn^{4+}$ ions to the elastic strains brings about the softening of the elastic constant $C_{44}$ in single crystal $Pr_{1-x}Ca_xMnO_3$ ($x$=0.35, 0.4, 0.5) [17], which also can lead to structural changes. The considerable softening of the ultrasound and the spontaneous lattice distortion in the vicinity of $T_{CO}$ for the $Pr_{1-x}Ca_xMnO_3$ ($0.5\leq x\leq 0.875$) series is, thus, resulted from both the couplings of the elastic strains to the quadrupole moment and to the charge-fluctuation mode of the $Mn^{3+}$ and $Mn^{4+}$ ions. Since the OO and CO in the present $Pr_{1-x}Ca_xMnO_3$ series occur simultaneously, it's difficult to clarify which is more important in the spontaneous lattice distortion. Judging from the magnitude of the softening of $(C_{11}-C_{12})/2$ (e.g. $\Delta C_{ij}/C_{ij} \approx -18\%$ for $Pr_{0.6}Ca_{0.4}MnO_3$) and $C_{44}$ ($\Delta C_{44}/C_{44} \approx -3.5\%$ for $Pr_{0.6}Ca_{0.4}MnO_3$) in the vicinity of $T_{CO}$ for $Pr_{1-x}Ca_xMnO_3$ ($x$=0.35, 0.4, 0.5) [17,18], it is suggested that the coupling of the quadrupole moment to the elastic strain is more important than the coupling of the charge-fluctuation mode of the $Mn^{3+}$ and $Mn^{4+}$ ions to the elastic strains in the spontaneous lattice distortion. To further clarify this point, it's very necessary to make ultrasonic measurements on single crystal $Pr_{1-x}Ca_xMnO_3$ ($0.5 \le x \le 0.875$) series for symmetry analysis.

Although the lattice parameters for all compounds undergo distinct changes near $T_{CO}$, the splitting of the Bragg peaks in the CO state (e.g. 40 K) is different for $x \le 0.75$ and $x > 0.75$ (see Fig. 6), which implies different configuration of the lattice parameters. As seen in Fig. 8, there is $b/\sqrt{2}$ <a≈c with $(b/\sqrt{2})/a < 1$ for $x \le 0.75$, indicating that the crystal structure for $x \le 0.75$ is tetragonally compressed along the *b*-axis. However, the *a* and *c* axes show anomalous shrink with tremendous expansion in the *b* axis between $x$=0.75 and $x$=0.8 (Fig 8(a)), which, as a result, lead to an abruptly increase in $(b/\sqrt{2})/a$ which is larger than 1 for $x \ge 0.8$ (Fig 8(b)), suggesting a tetragonal elongation of the crystal structure along the *b*-axis with respect to ideal cubic lattice. Although the crystal structure undergoes large change between $x$=0.75 and $x$=0.8, the orthorhombic unit cell volume at 40 K decreases smoothly with increasing $x$ from 0.5 to 0.875, as seen in the inset of Fig. 6.

To get more detailed structural information, we show the Mn-O distances at 40K for $x$=0.625 and 0.8 in Fig. 6(a) and (b), respectively. The $MnO_6$ octahedral at 40 K for $x$=0.625 can be seen with three different Mn-O distances that is Mn-$O_a$=1.8576 Å along the *b*-axis, and Mn-$O_p$ =1.9358 Å and 2.0058 Å in the *a*-*c* plane, implying a JT distortion of the apically compressed type (i.e. the $Q_2$-mode JT distortion). In contrast, the four short Mn-$O_p$ distances in the *a*-*c* plane at 40 K for *x*=0.8 is almost comparable (Mn-$O_p$=1.9116 Å and 1.8907 Å), but much shorter than the two Mn-$O_a$ distance (1.9605 Å) along the *b*-axis, which shows an apically elongated type JT distortion (i.e. the $Q_3$-mode JT distortion). From the different types of the distortion of the $MnO_6$ octahedra, it is easy to understand that the tetragonally compressed and elongated crystal structures for $x \leq 0.75$ and $x > 0.75$ are mainly due to the $Q_2$- and $Q_3$-mode cooperative JT distortion, respectively, and it is the crossover of the JT vibration mode from $Q_2$ to $Q_3$ that induces the change of the crystal structure between *x*=0.75 and *x*=0.8. Since the orbital ordering, and hence the magnetic structures is closely related to different types of cooperative JT vibration modes [28], the apically compressed $MnO_6$ octahedra for $x \leq 0.75$ implies a $3d_{x^2-r^2}/3d_{y^2-r^2}$ polarization of the $e_g$ orbitals, while the apically elongated octahedral for $x > 0.75$ implies a $3d_{z^2-r^2}$ polarization of the $e_g$ orbitals, which is consistent with the neutron diffraction results [27,33]. These effects of the cooperative JT effect on the crystal and magnetic structures, together with its effect on the stability of the CO state for $R_{1-x}Ca_xMnO_3$ (R=La [30], Pr), strongly suggests that the cooperative JT effect is one of the most important ingredients that governs the physics of the charge ordered state for perovskite type manganites.

## IV. Conclusions

We have studied the transport, structural and ultrasound properties for $Pr_{1-x}Ca_xMnO_3$ ($0.5 \leq x \leq 0.875$). The electronic conduction in the CO state for $Pr_{1-x}Ca_xMnO_3$ series can be best described by the VRH mechanism. The magnetic field induced deviation of the electronic conduction from VRH can be understood within the framework of the phase separation model. The strong correlation between the magnetoresistance effect and the cooperative JT effect strongly suggests the cooperative JT lattice distortion and resultant electron-lattice interaction is one of the key factor that determines the stability of the CO state. The powder X-ray diffraction revealed a crossover of the JT vibration mode from $Q_2$ to $Q_3$ between *x*=0.75 and *x*=0.8, which induces a change of the crystal structure from *b*-axis compressed orthorhombic to *b*-axis elongated one, and orbital ordering from $3d_{x^2-r^2}/3d_{y^2-r^2}$ type to $3d_{z^2-r^2}$ type. The results suggest that the cooperative JT effect plays an essential role of physics in the CO state for the $Pr_{1-x}Ca_xMnO_3$ system.

This work was supported by the Chinese National Nature Science Fund, and the Ministry of Science and Technology of China.

[1] S. Jin, T. H.Tiefel, M. McCormack, R. A. Fastnacht, R. Ramesh, and L. H. Chen, Science **264**, 413 (1994).

[2] S. Mori, C. H. Chen, and S-W. Cheong, Nature (London) **392**, 473 (1998).

[3] P. G. Radaelli, D. E. Cox, L. Capogna, S.-W. Cheong, M. Marezio, Phys. Rev. B **59**, 14440 (1999).

[4] C. Martin, A. Maignan, M. Hervieu, and B. Raveau, Phys. Rev. B **60**, 12191 (1999).

[5] Y. Tomioka, A. Asamitsu, H. Kuwahara, Y. Moritomo, and Y. Tokura, Phys. Rev. B **53**, R1689 (1996).

[6] Y. Moritomo, H. Kuwahara, Y. Tomioka, Y. Tokura, Phys. Rev. B **55**, 7549 (1997).

[7] A. Asamitsu, Y. Tomioka, H. Kuwahara, and Y. Tokura, Nature (London) **388**, 50 (1997).

[8] J. Stankiewicz, J. Sese, J. Garcia, J. Blasco, C. Rillo, Phys. Rev. B **61**, 11236 (2000).

[9] V. Kiryukhin, D. Casa, J. P. Hill, B. Keimer, A. Vigliante, Y. Tomioka, and Y. Tokura, Nature (London) **386**, 813 (1997).

[10] N. A. Babushkina, L. M. Belova, O. Yu. Gorbenko, A. R. Kaul, A. A. Bosak, V. I. Ozhogin, and K. I. Kugel, Nature (London) **391**, 159 (1998).

[11] Y. Tomioka, A. Asamitsu, H. Kuwahara, and Y. Tokura, J. Phys. Soc. Jpn. **66**, 302 (1997).

[12] S. Hébert, A. Maignan, R. Frésard, M. Hervieu, R. Retoux, C. Martin, B. Raveau, Eur. Phys. J. B **24**, 85 (2001).

[13] X. G. Li, R. K. Zheng, G. Li, H. D. Zhou, R. X. Huang, J. Q. Xie, and Z. D. Wang, Europhys. Lett. **60**, 670 (2002).

[14] R. K. Zheng, R. X. huang, A. N. Tang, G. Li, X. G. Li, J. N. Wei, J. P. Shui, Z. Yao, Appl. Phys. Lett. **81**, 3834 (2002).

[15] M. von Zimmermann, C. S. Nelson, J. P. Hill, D. Gibbs, M. Blume, D. Casa, B. Keimer, Y. Murakami, C. C. Kao, C. Venkataraman, T. Gog, Y. Tomioka, Y. Tokura, Phys. Rev. B **64**, 195133 (2001).

[16] V. S. Shakhmatov, N. M. Plakida, N. S. Tonchev, JETP Lett+ **77**, 15 (2003).

[17] H. Hazama, Y. Nemoto, T. Goto, Y. Tomioka, A. Asamitsu, and Y. Tokura, Phys. B **312-313**, 757 (2002).

[18] T. Goto and B Lüthi, Adv. in Phys. **52**, 67 (2003).

[19] H. Fujishiro, T. Fukase, M. Ikebe, J. Phys. Soc. Jpn. **70,** 628 (2001).

[20] A. P. Ramirez, P. Schiffer, S-W. Cheong, C. H. Chen, W. Bao, T. T. M. Palstra, P. L. Gammel, D. J. Bishop, and B. Zegarski, Phys. Rev. Lett. **76**, 3188 (1996).

[21] K. Vijaya Sarathy, S. Parashar, A. R. Raju, and C. N. R. Rao, Solid State Sci. **4**, 353 (2002).

[22] N. F. Mott and E. A. Davis, *Electronic Processes in Non-Crystalline solids*, 2nd ed. (Oxford University Press, New York, 1979).

[23] S. Mori, C. H. Chen, and S-W. Cheong, Phys. Rev. Lett. **81**, 3972 (1998).

[24] M. Mayr, A. Moreo, J. A. Vergés, J. Arispe, A. Feiguin, and E. Dagotto, Phys. Rev. Lett. **86**, 135 (2001).

[25] M. Roy, J. F. Mitchell, A. P. Ramirez, and P. Schiffer, J. Phys.: Condens. Matter **11**, 4843 (1999).

[26] E. O. Wollan, W. C. Koehler, Phys. Rev. **100**, 545 (1955).

[27] Z. Jirák, S. Krupička, Z. Šimša, M. Dlouhá, and S. Vratislav, J. Magn. Mag. Mater. **53**, 153 (1985).

[28] T. Hotta, A. L. Malvezzi, E. Dagotto, Phys. Rev. B **62**, 9432 (2000).

[29] R. D. Shannon, Acta Cryst. **A32**, 751 (1976).

[30] R. K. Zheng, G. Li, A. N. Tang, Y. Yang, W. Wang, X. G. Li, Z. D. Wang, H. C. Ku, Appl. Phys. Lett. **83**, xxxx (2003) (in press).

[31] R. L. Melcher, in *Physical Acoustics*, Vol. 12, Edited by W. P. Mason and R. N. Thurston, (Academic Press, New York) 1976, p.1.

[32] R. A. Young, A. Sakthivel, T. S. Moss, and C. O. Paiva-Santos, J. Appl. Crystallogr. **28**, 366 (1994).

[33] Z. Jirák, F. Damay, M. Hervieu, C. Martin, B. Raveau, G. André, and F. Bourée, Phys. Rev. B **61**, 1181 (2000).

# Figure captions

Fig. 1. Temperature dependence of the resistivity under magnetic fields up to 14T for $Pr_{1-x}Ca_xMnO_3$ ($0.5\leq x\leq 0.875$). For clarity, the resistivity for $x$=0.6 was magnified by 5 folds.

Fig. 2(a) The ln$\rho$ versus $T^{-1/4}$ curves under fields up to 9T for $Pr_{0.5}Ca_{0.5}MnO_3$. Inset (a) in Fig.2(a) is the $\rho$-$T$ curve for $H$=14T, the solid line is the fit according to Eq.(1). Inset (b) in Fig.2(a) is the field dependence of the $T_{VRH}$ . Fig. 2(b) The ln$\rho$ versus $T^{-1/4}$ curves under fields up to 10T for $La_{0.5}Ca_{0.5}MnO_3$. Inset (a) in Fig. 2(b) is the $\rho$-$T$ curve for $H$=12T, 13T, and 14T, the solid lines are the fit according to Eq.(1). Inset (b) in Fig. 2(b) is the ln$\rho$ versus $T^{-1/4}$ curves for H=8T and 10T.

Fig. 3. The ln$\rho$ versus $T^{-1/4}$ curves at zero field for $Pr_{1-x}Ca_xMnO_3$ ($0.55\leq x\leq 0.875$).

Fig. 4. Temperature dependence of the relative change of the longitudinal ultrasound ($\Delta V/V$) for $Pr_{1-x}Ca_xMnO_3$ ($0.5\leq x\leq 0.875$).

Fig. 5. The relative change of the ultrasound ($\Delta V/V$) (open square) and the magnetoresistance effect (*MR*) (filled circle) at $H$=14T and $T$=90K as a function of the Ca concentration $x$.

Fig. 6. Selected X-ray Rietveld refinements of (a) $Pr_{0.375}Ca_{0.625}MnO_3$ and (b) $Pr_{0.2}Ca_{0.8}MnO_3$ at 40 K. The observed data are represented as points, the calculated data as a line, and the difference curve is plotted below. The positions of the Bragg reflections are marked. The reliability of the refinement are $R_P$ =10.54%, $R_{WP}$ =15.43% and $\chi^2$=2.33 for $Pr_{0.375}Ca_{0.625}MnO_3$, and $R_P$ =10.85%, $R_{WP}$=16.41% and $\chi^2$=2.59 for $Pr_{0.2}Ca_{0.8}MnO_3$. Inset in (a) is the unit cell volume of the orthorhombic

lattice at 40 K as a function of the Ca concentration $x$. The Mn-O distances at 40 K for $x$=0.625 and $x$=0.8 are also shown in Fig. 6(a) and (b), respectively.

Fig. 7. Temperature dependence of the lattice parameters for $Pr_{1-x}Ca_xMnO_3$ ($x$=0.625, 0.67, 0.75, 0.825). The solid lines are guides to the eyes.

Fig. 8. (a) The lattice parameters $a$, $b/\sqrt{2}$, and $c$ at 40 K as a function of the Ca concentration $x$. (b) The ratio of the lattice parameters $(b/\sqrt{2})/a$ as a function of the Ca concentration $x$. Insets in (b) are schematic illustrations of the tetragonally compressed and elongated orthorhombic structures with $Q_2$- and $Q_3$-mode JT distortion of the $MnO_6$ octahedra.

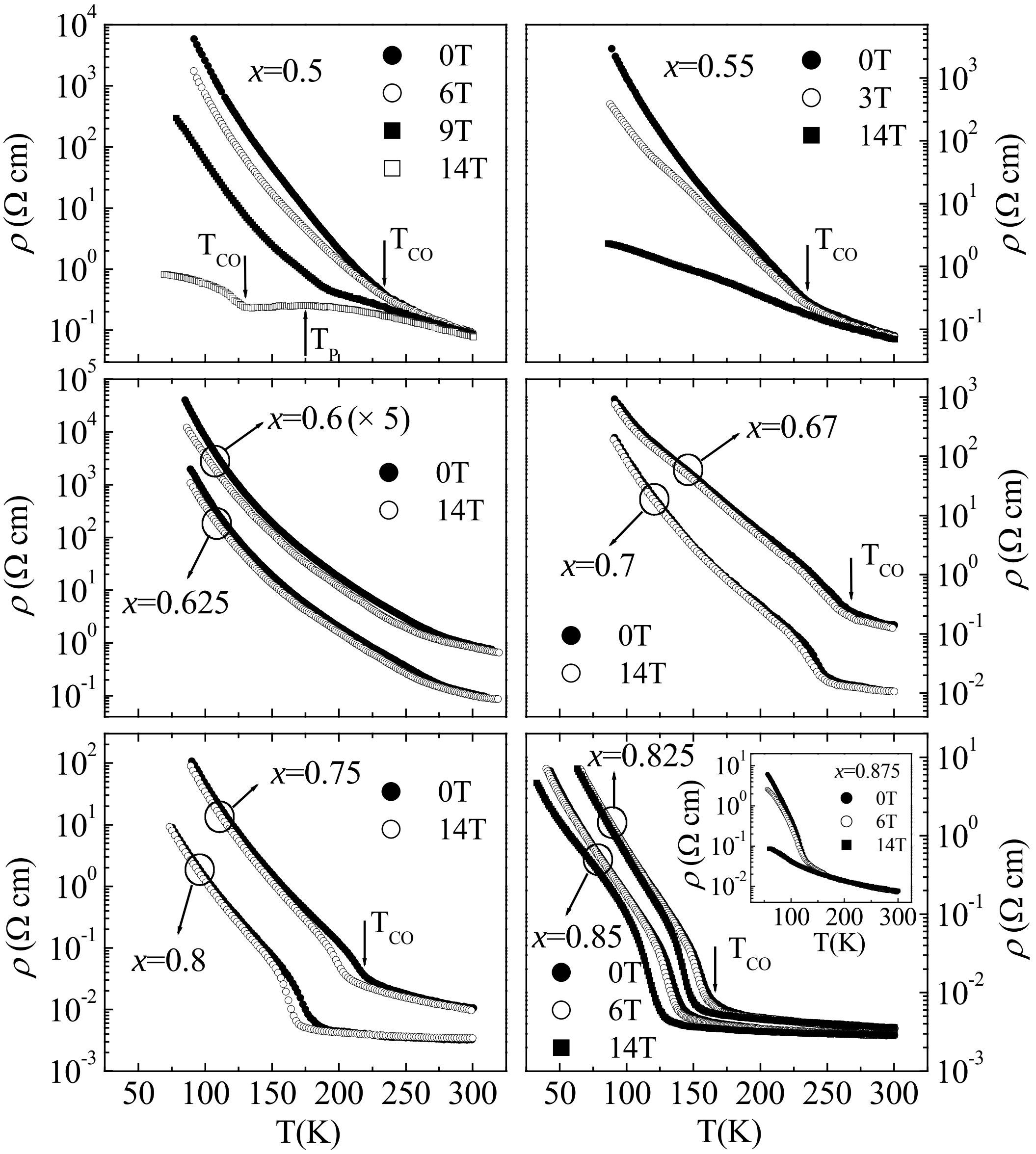

**Fig. 1 By R. K. Zheng *et al.***

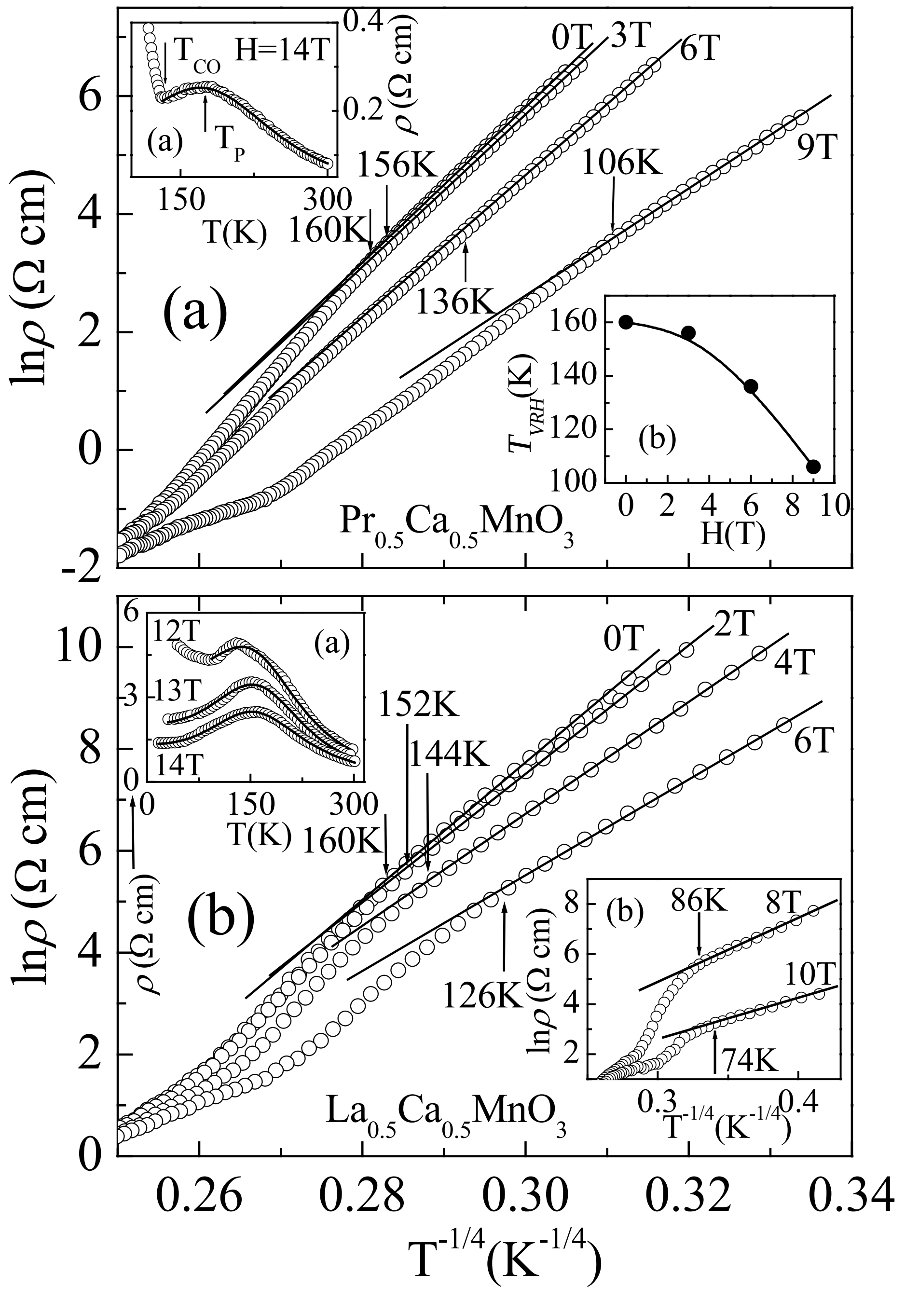

**Fig. 2 By R. K. Zheng *et al.***

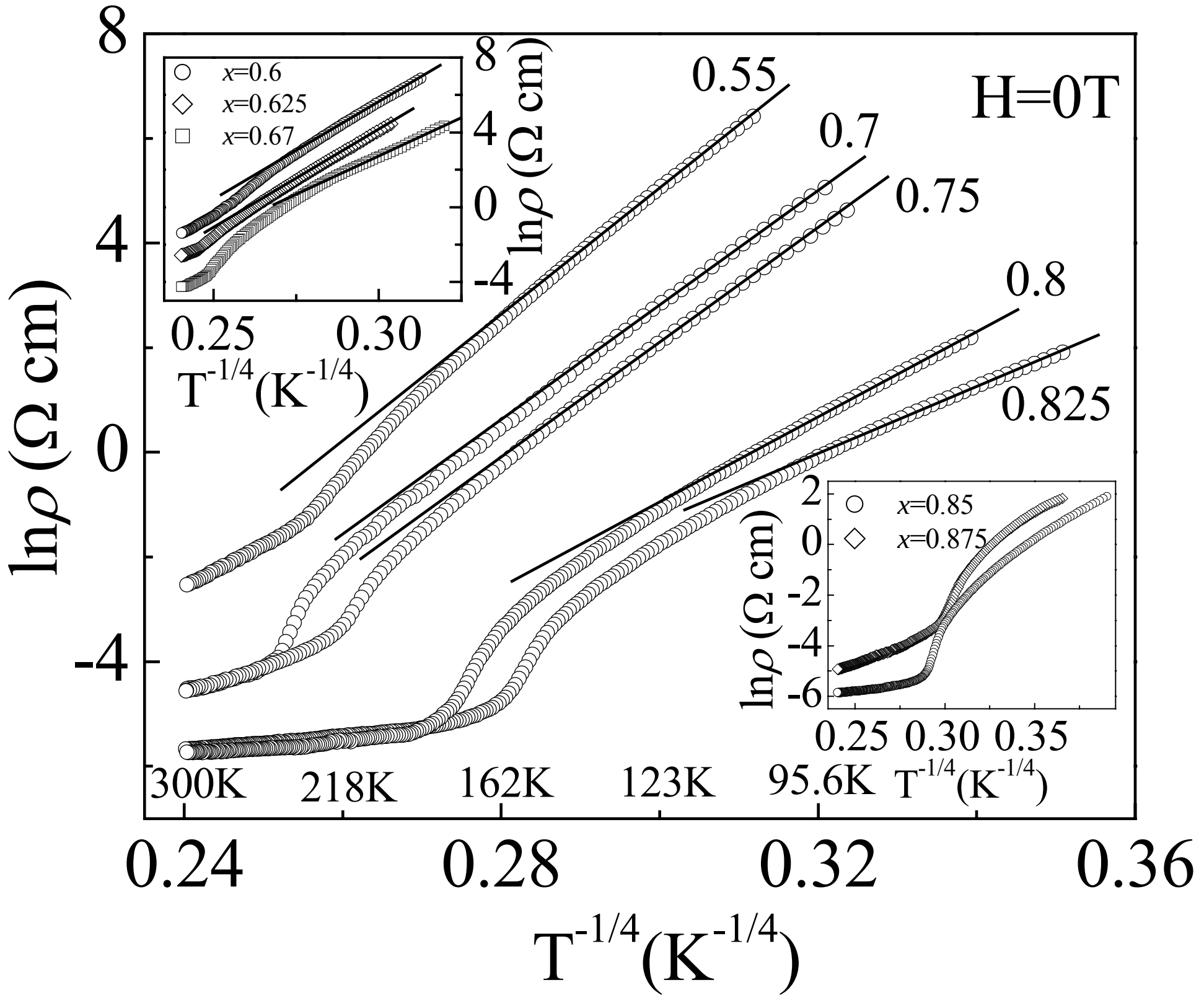

**Fig. 3 By R. K. Zheng *et al.***

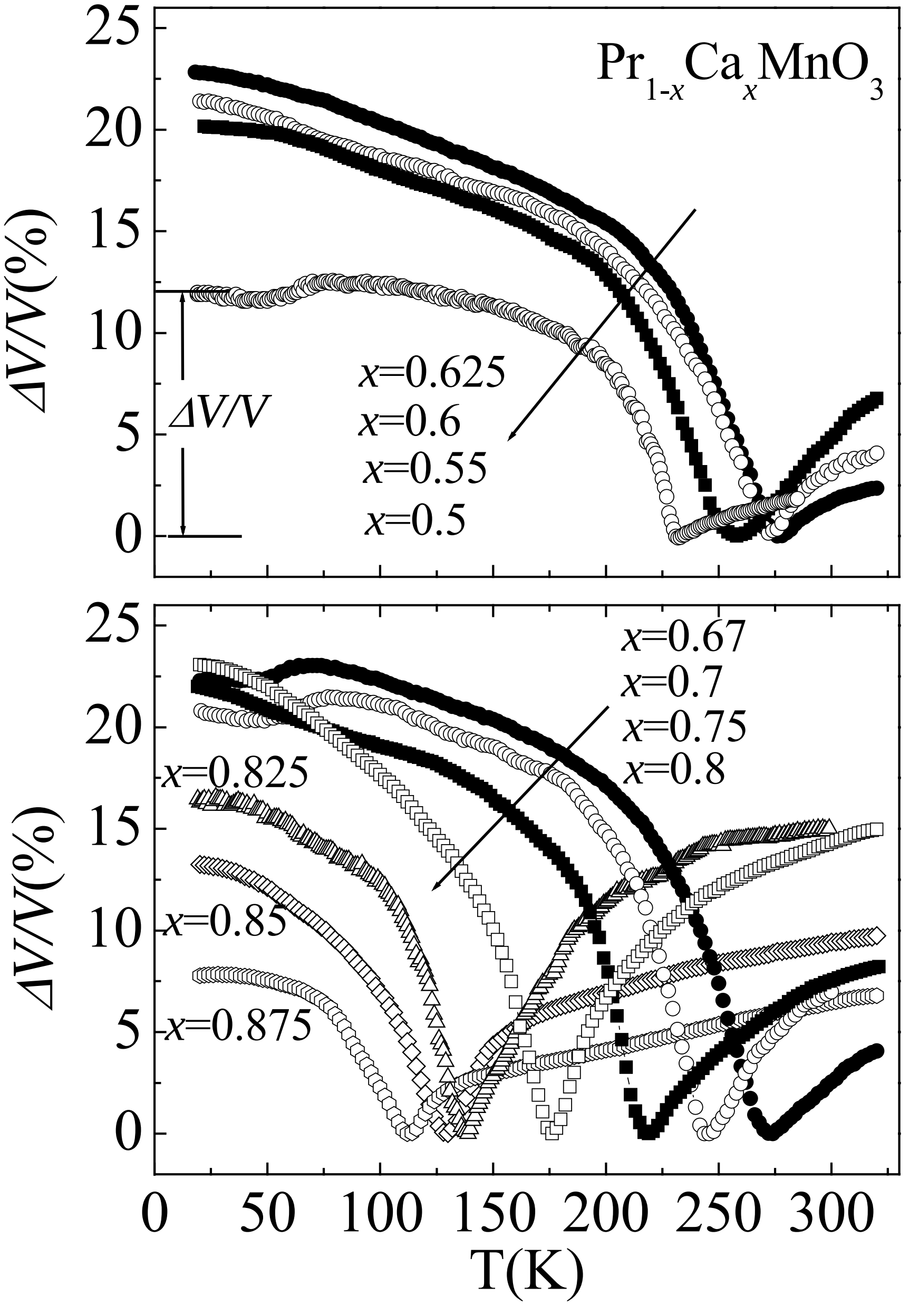

**Fig. 4 By R. K. Zheng *et al.***

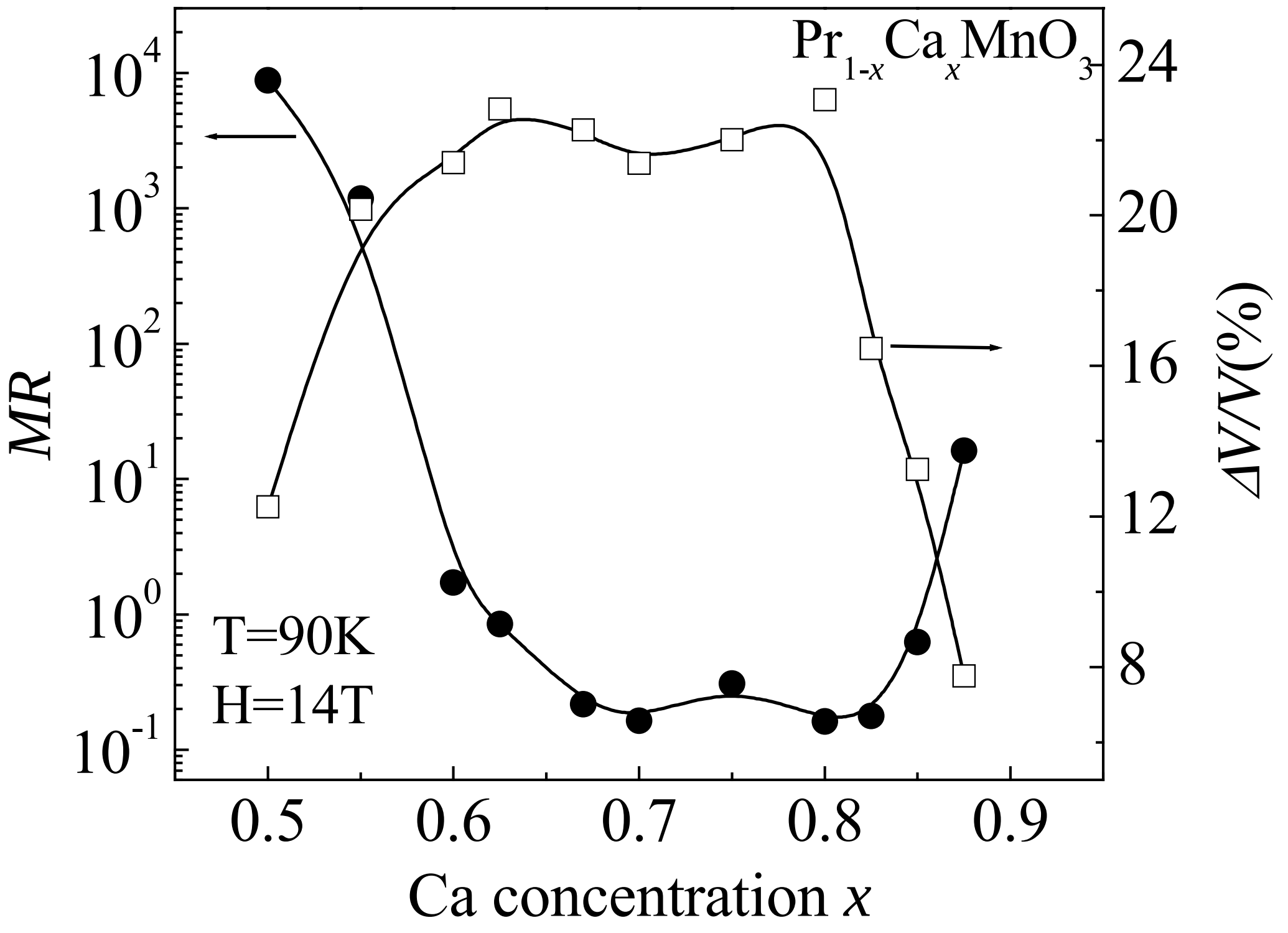

**Fig. 5 By R. K. Zheng *et al.***

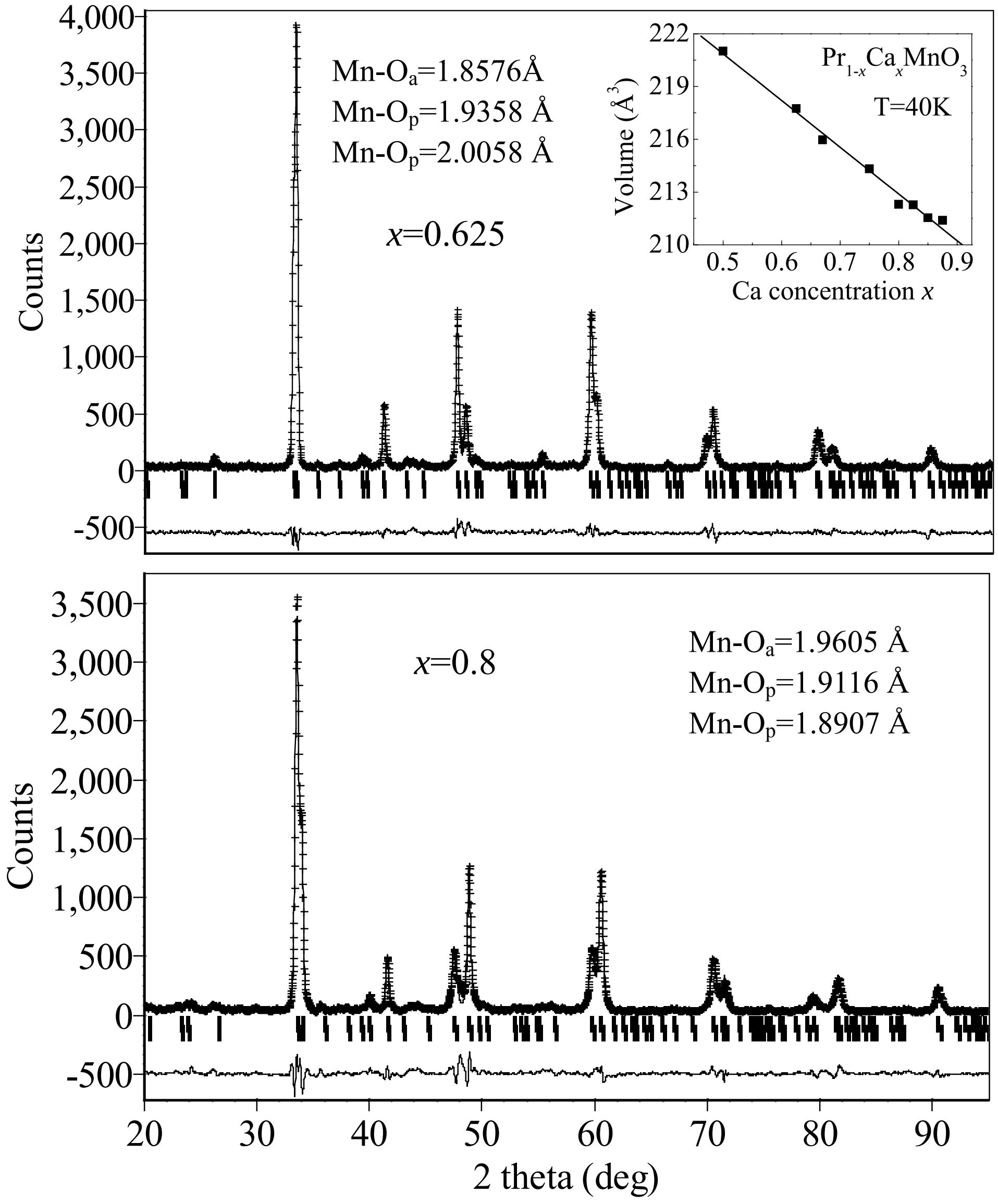

**Fig. 6 By R. K. Zheng *et al.***

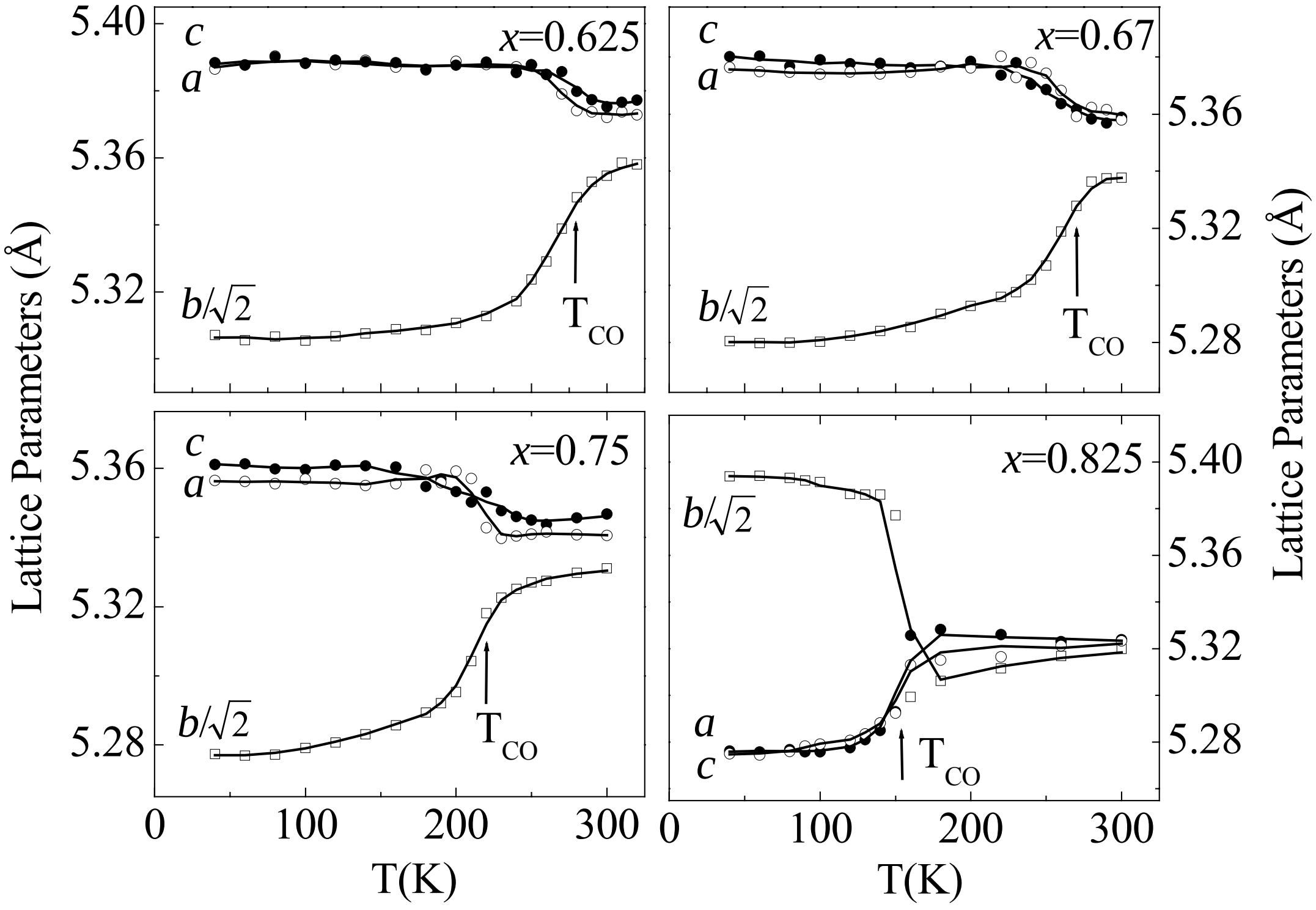

**Fig. 7 By R. K. Zheng *et al.***

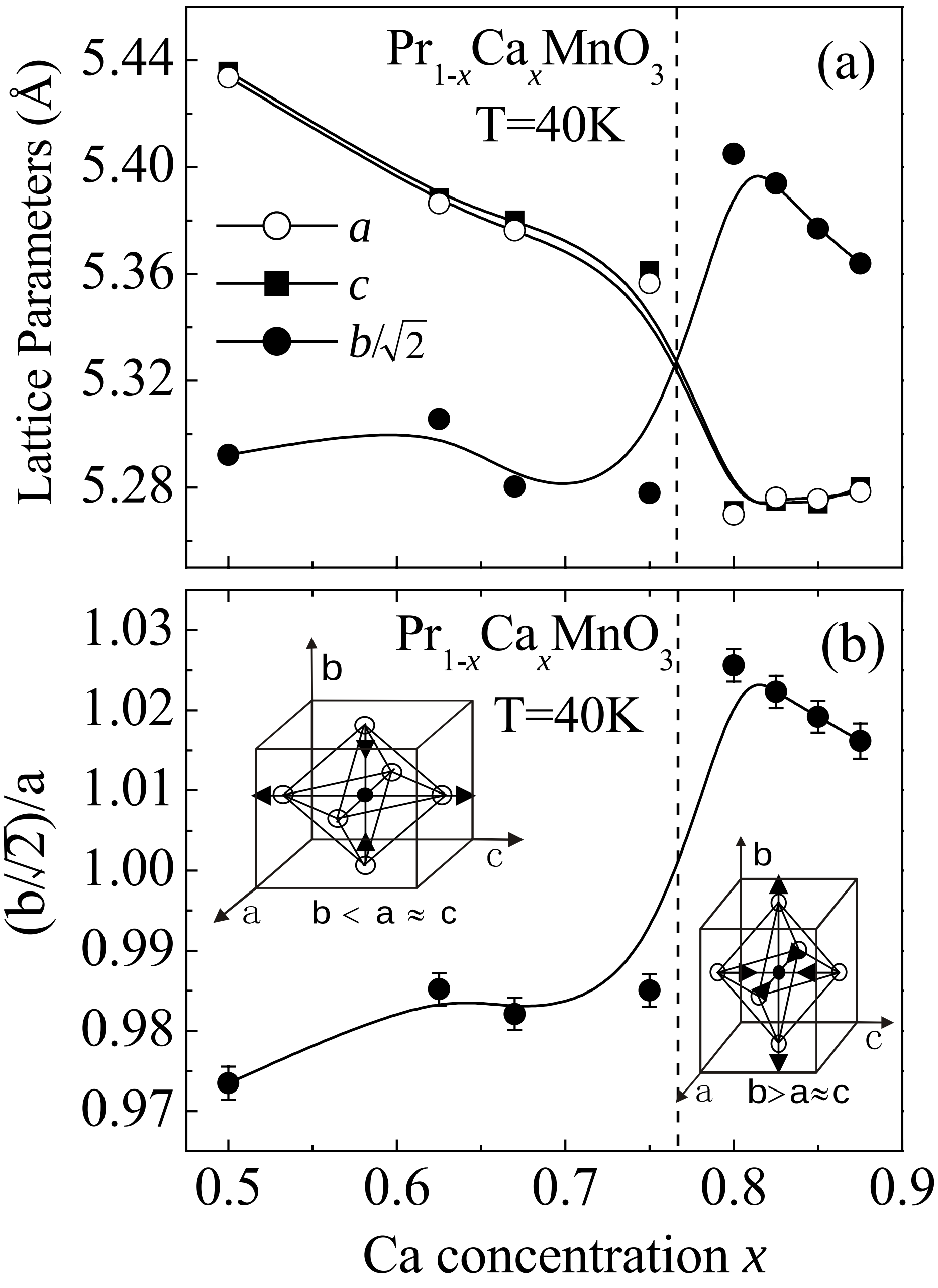

**Fig. 8 By R. K. Zheng *et al.***